# NoM: Network-on-Memory for Inter-bank Data Transfer in Highly-banked Memories

Seyyed Hossein SeyyedAghaei Rezaei[1]   Mehdi Modarressi[1,3]   Rachata Ausavarungnirun[2]
Mohammad Sadrosadati[3]   Onur Mutlu[4]   Masoud Daneshtalab[5]

[1]University of Tehran   [2]King Mongkut's University of Technology North Bangkok   [3]Institute for Research in Fundamental Sciences
[4]ETH Zürich   [5]Mälardalens University

**Abstract**— Data copy is a widely-used memory operation in many programs and operating system services. In conventional computers, data copy is often carried out by two separate read and write transactions that pass data back and forth between the DRAM chip and the processor chip. Some prior mechanisms propose to avoid this unnecessary data movement by using the shared internal bus in the DRAM chip to directly copy data within the DRAM chip (e.g., between two DRAM banks). While these methods exhibit superior performance compared to conventional techniques, data copy across different DRAM banks is still greatly slower than data copy within the same DRAM bank. Hence, these techniques have limited benefit for the emerging 3D-stacked memories (e.g., HMC and HBM) that contain hundreds of DRAM banks across multiple memory controllers. In this paper, we present Network-on-Memory (NoM), a lightweight inter-bank data communication scheme that enables direct data copy across both memory banks of a 3D-stacked memory. NoM adopts a TDM-based circuit-switching design, where circuit setup is done by the memory controller. Compared to state-of-the-art approaches, NoM enables both fast data copy between multiple DRAM banks and concurrent data transfer operations. Our evaluation shows that NoM improves the performance of data-intensive workloads by 3.8X and 75%, on average, compared to the baseline conventional 3D-stacked DRAM architecture and state-of-the-art techniques, respectively.

## 1 INTRODUCTION

THE memory subsystem is a major key performance bottleneck and energy consumer in modern computer systems. As a major challenge, the off-chip memory bandwidth does not grow as fast as the processor's computation throughput. The limited memory bandwidth is the result of relatively slow DRAM technologies (necessary to guarantee low leakage), the non-scalable pinout, and on-board wires that connect the processor and memory chips.

Prior work reports that a considerable portion of the memory bandwidth demand in many programs and operating system routines is due to bulk data copy and initialization operations [1]. Even though moving data blocks that already exist in memory from one location to another location does not involve computation on the processor side, the processor has to issue a series of read and write operations that move data back and forth between the processor and the main memory.

Previous works address this problem by providing in-memory data copy operations [1-5]. RowClone [1] provides a mechanism for both intra-bank and inter-bank copy, but RowClone's main focus is to enable fast data copy inside a DRAM subarray by moving data from one row to another through the row buffer inside the same DRAM subarray [17]. As we show in Section 3, several programs and operating system services copy data *across DRAM banks*. In this case, RowClone uses the internal bus that is shared across all DRAM banks and moves data one cache block at a time. During the copy, the shared internal DRAM bus is reserved and other memory requests to the DRAM chip are therefore delayed.

Emerging 3D-stacked DRAM architectures integrate hundreds of banks in a multi-layer structure. Data copy in 3D-stacked DRAM is much more likely to occur *across* different DRAM banks. RowClone does not provide as high benefits as for intra-bank copy when copying data across banks since it leverages the low-bandwidth shared bus between banks to move data. Aside from the overhead of cross-bank data copies, DRAM banks of the 3D memories are partitioned into several sets, each set having its own independent memory controller. RowClone's design does not provide for data movement between *independently-controlled* banks. While LISA [3] improves RowClone's performance by supporting faster inter-subarray copies, it does not provide any improvement for inter-bank copies.

To allow direct data copy between memory banks in 3D-stacked DRAM, we propose to interconnect the banks of a highly-banked 3D-stacked DRAM using a lightweight network-on-memory (NoM). NoM carries out copy operations entirely within the memory across DRAM banks, without any intervention from the processor. To copy data across different banks, NoM adds a Time-division Multiplexed (TDM) circuit-switching mechanism to the 3D-stacked DRAM memory controller. With NoM, banks in 3D-stacked memory are equipped with very simple circuit-switched routers and rely on a central node to set up circuits. This centralized scheme is compatible with the current architecture of 3D memories because there already is a front-end controller unit that forwards all requests to their destination banks.

In addition to its compatibility with the current structure of 3D memories to improve speed of inter-bank data transfers, NoM's second main advantage over prior in-memory data transfer architectures [1-3] is its ability to perform *multiple* data transfers *in parallel*. To perform data transfers in parallel, NoM replaces the global links of the shared bus with a set of shorter inter-bank links and, hence, yields higher throughput and scalability. Although there are some proposals to interconnect multiple 3D memory cubes (chips) and processors via a network [6], to the best of our knowledge, NoM is the first attempt to implement a network across the banks of a 3D memory chip.

Our experimental results show that NoM outperforms conventional 3D DRAM architecture and RowClone by 3.8x and 75%, respectively, under the evaluated copy-intensive workloads.

## 2 NETWORK-ON-MEMORY

**Baseline 3D DRAM architecture**. Although NoM can be used in both traditional DRAM and 3D-stacked DRAM architectures, we specifically tailor our design for the emerging 3D-stacked memories, like the Hybrid Memory Cube (HMC) [7]. In the HMC architecture, up to 8 DRAM layers (for a total of 8GB capacity) are stacked on top of one logic layer. Each layer (either logic or DRAM) is divided into 32 slices. Each DRAM slice contains two DRAM banks. The logic and DRAM slices that are vertically adjacent form a column of stacked slices, called a vault [7]. Each vault has its own vault controller implemented on its logic die. Vaults can be accessed simultaneously. Each vault consists of several (e.g., 4-8) banks that share a single set of internal data, address, and control buses. Each bank contains tens of (e.g., 64-128) subarrays, each of which consists of hundreds of (e.g., 512-2048) rows of DRAM cells that share a global row buffer that enables access to data in a given row.

**NoM architecture**. In addition to the conventional address, data, and control buses, NoM connects each bank to its neighboring banks at the X, Y and Z dimensions to form a 3D mesh topology. We use the mesh topology due to its simple structure and short non-intersecting links, which do not significantly change the structure of the DRAM layer.

We show the high-level organization of a 2D DRAM chip (for simplicity) with NoM in Fig. 1a. The dashed lines represent the extra links added to establish a network. NoM adopts TDM-based circuit-switching [8]. In a TDM network, each circuit reserves one or multiple time slots in a repeating time window. With n-cycle time windows, each router has an n-entry slot table, which is filled by the circuit control unit during circuit setup. This table determines the input-output connections of the router at each cycle of the time window (see Section 2.1). To manage this circuit switch, NoM introduces a centralized circuit control unit (CCU). The CCU receives internal data transfer commands from the processor and finds and reserves a circuit, as a sequence of TDM time slots, from the source to the destination banks.

In most 3D-stacked memories, including the HMC [7], there are two levels of logic between the processor and memory banks: (1) the front-end controller of the memory chip that directs processor requests to the corresponding vault via a crossbar and (2) vault controllers, each of which acts as the memory controller for the banks within the vault. Since the front-end controller has a global view of all memory transactions, the CCU can be added directly to the front-end controller.

Fig. 1b shows the architecture of a bank in more detail. The components unique to NoM are colored in yellow. The bank, as the figure indicates, can send/receive data to/from either 1) the network through the network links/circuit switching buffers (CS Buf) for direct data transfer operations or 2) the conventional buses via the bank IO buffer for regular read/write operations. NoM's circuit-switched router is simple and only consists of a crossbar, a latch (associated with each network link to keep incoming data for a single cycle), and a local controller unit. This local controller unit (Ctrl) is simply a time-slot table that determines the input-output connections of NoM ports (via the internal crossbar of the bank) on a per-cycle basis. The time-slot table is programmed by the CCU. These components are all compatible with the current DRAM technology and can be integrated into off-the-shelf 3D DRAM chips (Note that current DRAM chips already have crossbars and latches). The circuit-switched router does not need the complex logic that a typical packet-switched router employs for buffering, routing, arbitration, hop-by-hop flow control, and VC allocation [8]. Thus, circuit-switching has smaller routers and is faster than packet-switching (e.g., it has one hop per cycle packet traversal).

### 2.1 TDM Slot Allocation

The NoM circuit setup is done in the CCU in a centralized manner. The CCU keeps the state of all reserved time slots across the network and services a new request by finding a sequence of time slots along one of the paths between the source and destination banks. The TDM slot allocation is a complex task, as it should guarantee collision-free movement of data.

Specifically, the allocation must guarantee that (1) no time slot of a link is shared by multiple circuits, and (2) a circuit must use increasingly numbered slots in consecutive routers to avoid data buffering. For example, if time slot *m* is allocated to a circuit in a router, time slot *m+1* should be allocated to it in the next router to have the data advance one hop per cycle without any need for buffering, arbitration, and hop-by-hop flow control.

NoM utilizes the fast and efficient centralized circuit setup mechanism presented in prior work [9]. NoM relies on a hardware accelerator to explore all the possible paths from the source

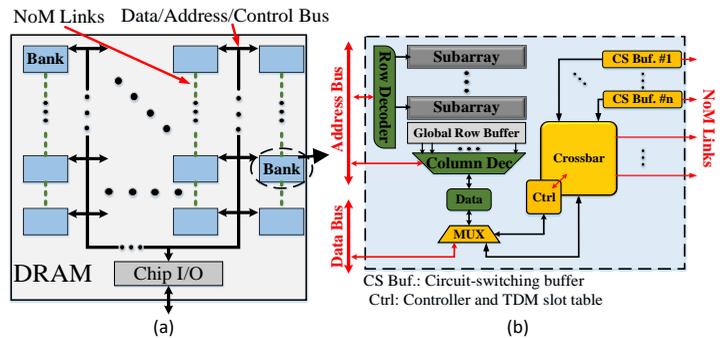

Fig. 1. (a) DRAM structure with NoM links, and (b) The modified architecture of a memory bank. The components that are unique to NoM are colored in yellow

to the destination in parallel. This hardware accelerator is composed of a matrix of simple processing elements (PEs), each associated with a network node. Assuming that the circuit reservation is done on *n*-slot windows and each router has *p* output ports, each accelerator PE keeps the occupancy state of its corresponding network node in a 2-D matrix $V$ of size $p \times n$. $V[i][j]=1$ indicates that the j$^{th}$ slot of the i$^{th}$ output port of the corresponding network router is reserved. To find a path, the source PE (associated with the source node of the circuit) uses an *n*-bit vector with all elements initialized to zero. This bit vector keeps track of the empty time slots of different paths and is propagated through all the shortest paths to the destination. If time slot x is free in this router, we need time slot x+1 to be free in the next router. Hence, in each PE, the bit vector is first rotated right and then ORed with the vectors corresponding to the output ports along the shortest path to eliminate (mark as unavailable) busy slots from the bit vector. It is then passed to the next PE towards the destination cell. Available circuit paths (sequence of time slots) appear as zero bits in the vector at the destination PE. The circuit path and time slots can be reserved by tracing back the path towards the source PE.

With *B*-bit links, *B* bits of data is transferred on NoM link in each cycle. If a circuit has *V* bits to transfer, the time-slots remain reserved for *V/B* time windows. The data transfer can be accelerated by reserving multiple slots, provided that the algorithm returns more than one free slot. After that, the algorithm is allowed to use the time-slot for the next requests.

### 2.2 Data Transfer on NoM

Fig. 2 shows the implementation of NoM on a 3D-stacked HMC-like DRAM architecture. The operation of the network is divided into two steps: circuit setup and data transfer over the circuit. Similar to prior work, we assume that the processor issues a specific direct data copy request that is separate from the regular read/write requests [1-5]. A direct data copy is handled by the CCU as follows:

1. The CCU queues and services the direct data copy requests in a FIFO manner. For each request, a circuit is established between the source and destination banks using the TDM slot allocation logic (See Section 2.1). Fig. 2 shows an example of this operation. In Fig. 2, there is a direct data copy request from bank A to bank B, arriving at the CCU at time t. Assume that NoM has 8-slot windows and the currently active time slot (based on which the router connections are configured) is time slot 0. Fig. 2 shows an example circuit between bank A and bank B that comprises of five consecutive time slots in five routers, starting from time slot 3 (as the starting slot of the earliest available circuit) at the router associated with bank A.

2. The CCU sends a read request to the source vault controller to generate the data read signals for the target block at the source bank A at time slot 3.

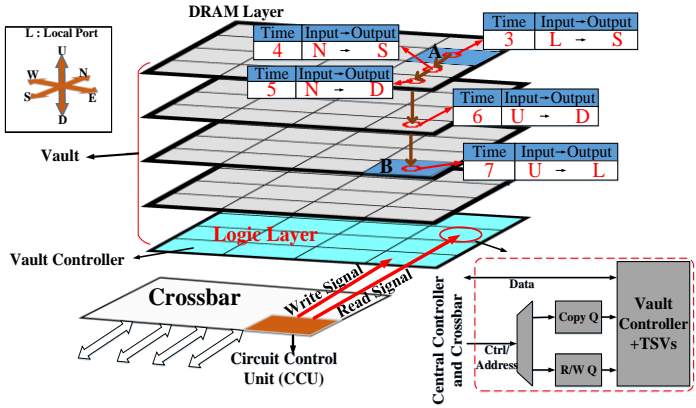

Fig. 2. NoM on 3D DRAM and an example circuit from bank A to bank B

3. The CCU waits for data to traverse the circuit and reach the destination. CCU knows when the data arrives at the destination bank because circuit-switching has deterministic transmission time.

4. Finally, the CCU sends a write request to the destination vault controller to write the received block to the destination bank B at time slot 7.

Once a request is picked up at cycle t, the CCU takes three cycles to route the request (one cycle to find a path, one cycle to establish the circuit path by configuring the slot tables along the circuit path, and one cycle to issue the read request and make data ready). So, the earliest time that the algorithm can consider for the data transfer on the circuit to start is t+3. CCU updates the slot tables by using dedicated links.

In our design, all vault controllers can serve both regular memory accesses and DRAM refresh requests in parallel with NoM's data copy operations. The only exception is when the corresponding vault controllers are busy reading and writing NoM data at the source and destination banks, respectively. In between these read and write steps, the data block passes on the NoM links and leaves the read/write bandwidth idle for regular accesses. NoM's ability to service concurrent copy operations as well as other regular memory accesses is one of its main advantages over previous works [1, 3].

A vault controller, as shown at the bottom right of Fig. 2, stores regular requests received from the front-end memory controller in a regular queue (R/W Q) and read/write requests related to direct copy requests to a high-priority queue (Copy Q). Direct copy requests are distinguished with a flag set by CCU.

From the circuit-switched router's point of view, the entire process is very simple. The circuit-controller reads the slot table entry for each time slot, shown by the <t,I→O> tuples in Fig. 2, and connects the input-output (I-O) pair indicated by the tuple using the crossbar. At the source and destination banks, circuit-controller also sets up the multiplexer inside the bank (MUX in Fig. 1b) to exchange data with memory. For example, at time slot 7, the slot table connects port U to port L (local ejection port) to forward the data received from input port U to output port L, and from there, to the data buffer of the bank.

**Correctness**. We handle memory consistency and cache coherence similarly to previously proposed in-memory data copy designs [1][3]. To keep the memory system coherent, (1) the memory controller writes back the source block of a copy operation if it is modified in cache (or all modified blocks from the source region, in case of bulk copy) before a copy, and (2) invalidates all cached blocks of the destination region present in cache.

Memory consistency should be enforced by software, i.e. software should insert special synchronization instructions when it is necessary to enforce order or atomicity [1].

### 2.3 NoM Implementation

We implement NoM on an HMC-like 3D-stacked memory. Each bank is connected to up to 6 other banks along the three dimensions. The link width is set to the internal memory bus width, i.e. 64 bits. Short planar links are used to connect two adjacent banks in a layer. For the vertical dimension, HMC already has a set of vertical links, implemented with the Through Silicon Via (TSV) technology. A TSV carries address, data, and control signals between the vault controller and DRAM banks. NoM with a full 3D mesh topology needs some extra TSVs to implement 3D mesh links in the third dimension (which connect every pair of vertically adjacent banks). To reduce the area overhead of NoM's full 3D mesh, we also propose a low-overhead design called NoM-Light. This design eliminates the 3D mesh vertical links but shares the bandwidth of the already existing TSVs to perform data transfer vertically. This design is motivated by our observation that the probability of simultaneously using both existing TSVs and NoM's 3D mesh TSVs in a single cycle in NoM with full 3D mesh topology is 0.45% under low NoM load and 7.1% under high NoM load. This low conflict rate suggests that we can eliminate the full 3D mesh TSVs and use the conventional HMC TSVs also for transferring NoM data with no noticeable performance loss.

A TSV that carries the address/data signals of a vault is a bus (and not a point-to-point link as in a full 3D mesh), so NoM's data is transferred in a broadcast-based manner vertically [10]. The disadvantage of this NoM-Light design, compared to the NoM with a full 3D mesh, is that only a single data item can traverse the third dimension in each vault simultaneously. The advantage of the NoM-Light design, however, is that NoM data can traverse any number of hops in the third dimension in a single cycle. As the vertical links are very short, this single-cycle multi-hop traversal has no timing violation problem [11-12].

NoM uses special extra sideband TSVs to program the slot tables. In each cycle, CCU sets at most one slot table entry in each vault. All slot tables are connected to a shared vertical link. The link is 12 bits wide to set the right slot: 3 bits to select one of the 8 banks of the vault, 4 bits to select a slot in the 16-slot window, and 6 bits to carry the slot table data: 3 bits to select one of the six input ports and 3 bits to select one of the six output ports that should be connected at that slot.

## 3 EVALUATION

**Simulation environment.** We use *Ramulator* [13], a fast and extensible open-source DRAM simulator that can support different DRAM technologies, to evaluate the effectiveness of NoM. We measure system performance using instructions per cycle (IPC). Circuit-level parameters and memory timing parameters are set based on DDR3 DRAM [15]. The baseline target memory is a 4GB HMC-like architecture with 32 vaults, four DRAM layers, and two banks per DRAM slice (for a total of 256 banks). The NoM topology is an 8×8×4 mesh. The circuit-switching time window has 16 time-slots. All datapaths and links inside the memory are 64 bits wide. We integrate the intra-subarray and intra-bank direct data copy mechanisms of RowClone [1] and LISA [3] into NoM: this way, inter-bank data copy is carried by NoM, whereas intra-subarray/bank data copy is handled by RowClone/LISA. We compare NoM to two baselines, RowClone and 3D-stacked memory described above.

**Workloads.** We evaluate NoM on four different benchmarks: *fork*, one the most frequently-used system calls in operating systems, *fileCopy20*, *fileCopy40*, and *fileCopy60*, three copy-intensive benchmarks that model the *mcached* memory object caching system with different volumes of object copies [1]. Fig. 3 illustrates the breakdown of memory accesses of each benchmark. The memory access types are inter-bank and intra-bank copies, initialization, and regular read/write accesses from the program. For these benchmarks, 20% to 60% of the memory traffic is generated by inter-bank copy operations. Note that processor-intensive benchmarks, such as SPEC CPU, typically use a small number of page/data copies and hence, are not the target of NoM.

**Performance analysis.** Fig. 4 compares the performance (IPC) of the three evaluated memory configurations. We make three observations. First, NoM provides 75% higher IPC than RowClone on average, as it enables faster inter-bank copies and performs multiple inter-bank copies and other memory accesses concurrently. Second, increasing the data copy traffic rate between banks (by increasing the frequency of NoM links from 600 MHz to 1.25 GHz) provides more opportunity for NoM to exploit its inherently high throughput, increasing the IPC benefits of NoM over RowClone. Third, NoM-Light has only 5%-20% lower IPC than baseline NoM, because of its more limited bandwidth between layers. However, NoM-Light still greatly outperforms RowClone.

**Energy analysis.** To evaluate the energy consumption of NoM, we use the *DRAMPower* simulator [14] and 3D-stacked DRAM energy model from Micron [15]. Our experimental results show that our mechanism reduces energy per access by up to 3.2x in comparison with baseline DDR3 memory, as it eliminates costly data exchange with the processor for copy operations. In comparison to RowClone, NoM consumes up to 9% more energy, mainly due to the extra links and logic it uses.

**Area analysis.** NoM equips each 16 MB DRAM bank with a simple router. We evaluate the area of NoM-Light using the area models and parameters presented in [15] (for DRAM chips) and [16] (for TSVs). The results show that the area overhead of NoM (comprising buffers, slot table, crossbar, controller, links and TSVs) is negligible (below 1%) when compared to the area of a 16 MB bank of HMC.

**Operating frequency**. To traverse a network hop using circuit switching (which must be completed in a single cycle), the packet should go through input latch read, crossbar traversal, link traversal, and input latch write in the downstream router. We model these datapath elements of NoM routers using CACTI in 45nm DRAM technology to obtain the latency of each step. The results show that the total latency of a single hop is less than 300ps in 45nm technology. Hence, our simple NoM routers are implemented in the DRAM layer, and the routers can keep up with the speed of the logic layer, which is 1.25GHz. The critical path of the hardware accelerator that allocates TDM slots (described in Section 2.2) is below 500ps, so it can find a path in a single cycle.

To investigate how NoM's performance scales when NoM's link frequency is lower than DRAM frequency, we keep the logic layer frequency to 1.25GHz but reduce the frequency of NoM links by 25% and 50%. As Fig. 3 shows, NoM shows sublinear performance degradation when NoM's link frequency scales down. However, NoM still delivers better performance than RowClone because network latency comprises only part of the copy latency, while a major source of copy latency is the bank access latency. Furthermore, concurrent data transfers enabled by NoM can hide part of the link and bank latencies.

## 4  CONCLUSION

This paper introduces NoM, a new design for performing direct data copy between banks inside a 3D-stacked DRAM chip. The proposed method uses a network between DRAM banks in order to enable direct data transfer between them. NoM uses TDM

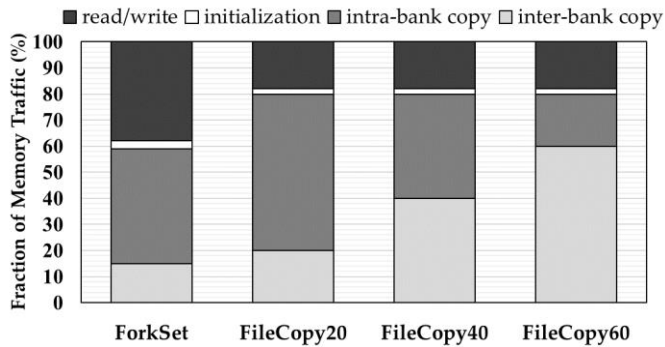

Fig. 3. Fraction of memory traffic due to different operations

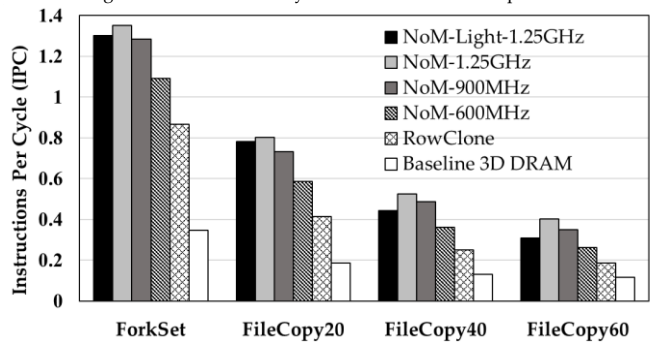

Fig. 4. Performance of NoM and NoM-Light versus baselines

-based circuit-switching design with a central control mechanism implemented in the memory controller. Our experimental results show an average of 75% and 3.8x performance improvement over the state-of-the-art (RowClone) and a conventional 3D DRAM architecture, respectively. NoM can be used in any highly-banked memory, including the 3D-stacked high bandwidth memory (HBM) architecture.

### REFERENCES

[1] V. Seshadri et al., "RowClone: Fast and energy-efficient in-DRAM bulk data copy and initialization," in *MICRO*, 2013.

[2] V. Seshadri, et al., "Ambit: in-memory accelerator for bulk bitwise operations using commodity DRAM technology," in *MICRO*, 2017.

[3] K. K. Chang, et al., "Low-Cost Inter-Linked Subarrays (LISA): Enabling fast inter-subarray data movement in DRAM," in *HPCA*, 2016.

[4] M. Gao et al., "DRAF: A Low-Power DRAM-Based Reconfigurable Acceleration Fabric," in *ISCA*, 2016.

[5] V. Seshadri *et al.*, "Fast Bulk Bitwise AND and OR in DRAM," in *IEEE CAL, 2015*.

[6] G. Kim et al., "Memory-centric system interconnect design with hybrid memory cubes," in *PACT*, 2013.

[7] Hybrid Memory Cube Specification 2.1, Nov. 2015, Hybrid Memory Cube Consortium, Tech. Rep.

[8]  T. Moscibroda et al.,"A Case for Bufferless Routing in On-chip Networks " in ISCA 2009.

[9] F. Pakdaman, A. Mazloumi, M. Modarressi, "Integrated Circuit-Packet Switching NoC with Efficient Circuit Setup Mechanism", in *Springer Journal of Supercomputing*, vol. 71, no. 8, 2015.

[10] D. Lee et al., "Simultaneous Multi-Layer Access: Improving 3D-Stacked Memory Bandwidth at Low Cost", in TACO 2016.

[11] S. H. Seyyedaghaei, et al., "Dynamic Resource Sharing for High-Performance 3-D Networks-on-Chip", in *IEEE CAL*, 2016.

[12] S. H. Seyyedaghaei, et al., "Fault-Tolerant 3-D Network-on-Chip Deign using Dynamic Link Sharing", in *DATE*, 2016.

[13] Y. Kim, et al., "Ramulator: A Fast and Extensible DRAM Simulator," in *IEEE CAL*, 2016.

[14] https://github.com/tukl-msd/DRAMPower

[15] Micron, DDR3 SDRAM system-power calculator, 2018.

[16] D. Kim, et al., "Neurocube: A Programmable Digital Neuromorphic Architecture with High-Density 3D Memory," in ISCA, 2016.

[17] Y. Kim et al., "A Case for Exploiting Subarray-Level Parallelism (SALP) in DRAM", in ISCA 2012.